# Spectral Formulation of the Elastodynamic Boundary Integral Equations for Bi-material Interfaces

K. Ranjith

SRM Research Institute, SRM University, Kattankulathur 603203, Tamil Nadu, India

E-mail: ranjith.k@res.srmuniv.ac.in, Phone: +91-44-27417906, Fax: +91-44-27456702

**Abstract**

A spectral formulation of the plane-strain boundary integral equations for an interface between dissimilar elastic solids is presented. The boundary integral equations can be written in two equivalent forms: (a) The tractions can be written as a space-time convolution of the displacement continuities at the interface (Budiansky and Rice, 1979) (b) The displacement discontinuities can be written as a space-time convolution of the tractions at the interface (Kostrov, 1966). Prior work on spectral formulation of the boundary integral equations has adopted the former as the starting point. The present work has for its basis the latter form based on a space-time convolution of the tractions. Tractions and displacement components are given a spectral representation in the spatial coordinate along the interface. The radiation damping term is then explicitly extracted to avoid singularities in the convolution kernels. With the spectral forms introduced, the space-time convolutions reduce to convolutions in time for each Fourier mode. Due to continuity of tractions at the interface, this leads to a simpler formulation and form of the convolution kernels in comparison to the formulation involving convolutions over the slip and opening history at the bi-material interface. The convolution kernels are validated by studying some model problems to which analytical solutions are known. When coupled with a cohesive law or a friction law at the interface, the formulation proposed here is of wide applicability for studying spontaneous rupture propagation.

1**. Introduction**:

The study of dynamic ruptures at bi-material interfaces has attracted much recent attention. This has been stimulated by interesting theoretical results that highlight the role of material asymmetry in dynamic rupture problems. Adams (1995) showed a short-wavelength instability in the response to perturbations from a state of steady sliding between dissimilar elastic solids with the Coulomb friction law acting at the interface. The mechanism of the instability is the destabilization of an interfacial elastic wave called the slip wave (Achenbach and Epstein, 1967). This suggests that the steady bi-material sliding problem with the Coulomb law is mathematically ill-posed and modifications to the Coulomb law are required. Experimentally motivated modifications to the Coulomb law have been discussed by Ranjith and Rice (2001) and Rice et al. (2001). Ranjith (2009) discovered a new instability of long-wavelength Love and Stoneley waves in slow sliding at a bi-material interface. Adams (1998) also obtained a slip pulse solution at a bi-material interface remotely stressed below the friction threshold. In that solution, a finite portion of the interface slips at any instant and the remainder of the interface is locked. The speed of propagation of the slip pulse is that of the slip wave. These results have prompted several numerical studies of dynamic ruptures at bi-material interfaces. In the literature, two types of numerical methods have been used to study bi-material interface rupture. Langer et al. (2013) and Kammer et al. (2014) have used finite element methods while Cochard and Rice (2000) and Ampuero and Ben-Zion (2008) have used the boundary integral equation method.

The latter two studies have for their basis the spectral formulation of the boundary integral equations for bi-material interfaces developed by Breitenfeld and Geubelle (1998). This method is especially suited to studying planar bi-material ruptures and has the advantage over the finite element method that it is less intensive computationally since field quantities only on the rupture plane need to be calculated unlike in the finite element method where field quantities away from the interface also need to be calculated. In the present paper, an alternative formulation to that of Breitenfeld and Geubelle (1998) for studying dynamic ruptures at bi-material interfaces is developed.

Consider a planar interface between two identical isotropic, homogeneous elastic half-spaces. A Cartesian coordinate system is located such that the interface is at $x_2 = 0$ and the $x_1$ coordinate is along the interface. It is assumed that the stresses and displacements are independent of the $x_3$ coordinate. Let $\tau_{ij}$ $(i, j = 1, 2)$ denote the in-plane stresses and $u_j$ $(j = 1, 2)$ denote the in-plane displacements. The traction components of stress at the interface are $\tau_j = \tau_{j2}$ $(j = 1, 2)$. The displacement discontinuities are given by

$$\delta_j(x_1, t) = u_j(x_1, x_2^{0+}, t) - u_j(x_1, x_2^{0-}, t) \ (j = 1, 2), \tag{1}$$

where $t$ denotes time.

Let $\lambda$ and $\mu$ be the Lame constants of the (identical) solids and let $c_s$ and $c_d$ be their shear and dilatational wave speeds, respectively. For an interface between identical solids, the boundary integral equations which relate the tractions and displacement discontinuities at the interface can be written in the form (Cochard and Madariaga, 1994):

$$\tau_1 = \tau_1^o - \frac{\mu}{2c_s}\frac{\partial \delta_1}{\partial t} + f_1(x_1,t),$$

$$\tau_2 = \tau_2^o - \frac{\lambda+2\mu}{2c_d}\frac{\partial \delta_2}{\partial t} + f_2(x_1,t),$$

(2)

where $\tau_j^o$ ($j=1,2$) are the tractions that will be present in the absence of displacement discontinuities at the interface and $f_j(x_1,t)$ ($j=1,2$) involve space-time convolutions over the displacement discontinuities. The second term on the right hand side of the two equations above is the radiation damping term that has been explicitly extracted to avoid singularities in the convolution kernels.

The spectral form of Eqn. (2) was introduced by Geubelle and Rice (1995). They wrote the tractions and displacement continuities in the form

$$\delta_j(x_1,t) = \sum_k D_j(k,t)\exp(ikx_1),$$

$$\tau_j(x_1,t) = \sum_k T_j(k,t)\exp(ikx_1),$$

(3)

where $k$ is the wavenumber and $D_j(k,t)$ and $T_j(k,t)$ are the amplitudes of the displacement discontinuities and the tractions, respectively. The convolution terms $f_j(x_1,t)$ ($j=1,2$) in Eqn. (2) then take the form

$$f_j(x_1,t) = \sum_k F_j(k,t)\exp(ikx_1),$$

(4)

where the functions $F_j(k,t)$ involve convolution over the time history of $D_j(k,t)$. The convolutions kernels for the case where the solids forming the interface are identical were derived by Geubelle and Rice (1995).

In the present work, the more general problem of an interface between dissimilar elastic solids is considered. Breitenfeld and Geubelle (1998) have made a prior generalization of the spectral formulation to the bi-material problem. In their approach, the elastodynamic convolutions are performed over the history of displacements or displacement discontinuities at the interface. Here, an alternative method is presented where the elastodynamic convolutions are performed over the history of tractions at the interface. Only results for 2D plane strain are reported here. Similar ideas can be applied to 2D anti-plane strain and subsequently be extended to a 3D formulation. Due to material dissimilarity across the interface, the rates of displacement discontinuity depend on both components of the tractions. In view of continuity of tractions at the interface, it becomes easier to obtain a relation between the rates of displacement discontinuity and the tractions as compared to the approach presented by Breitenfeld and Geubelle (1998). The convolution kernels for the bi-material problem are derived in this paper. The kernels are validated by numerical solution of some model problems where analytical solutions are known.

**2. Elastodynamic Relations**:

We derive the elastodynamic relations between the tractions and the slip and opening velocity at the interface between the two half-spaces. Field quantities and material properties relevant to the top half-space will be denoted by the superscript + while those relevant to the lower half-space will be denoted by the superscript -. In the development below, the superscripts will be dropped when it is clear from the context of the discussion which half-space is being referred to. First, consider the upper half-space. Let the tractions acting on the upper half-space at the plane $x_2 = 0^+$ be of the form

$$\tau_j = T_j(k,p)\exp(ikx_1 + pt) \ (j=1,2), \tag{5}$$

where $k$ is the wavenumber, $p$ is the Laplace variable and $T_j$ is the traction amplitude. The corresponding displacements are of the form

$$u_j = U_j(k,p)\exp(ikx_1 + pt) \ (j=1,2), \tag{6}$$

where $U_j$ denotes the displacement amplitude.

Following Geubelle and Rice (1995), the elastodynamic relations between the amplitudes of the tractions and the displacements can be written in the form

$$\begin{Bmatrix} T_1 \\ T_2 \end{Bmatrix} = \begin{bmatrix} G_{11} & G_{12} \\ G_{21} & G_{22} \end{bmatrix} \begin{Bmatrix} U_1 \\ U_2 \end{Bmatrix}, \tag{7}$$

where

$$\begin{aligned} G_{11}(k,p) &= -\mu |k| \frac{\alpha_d(1-\alpha_s^2)}{1-\alpha_s\alpha_d}, \\ G_{22}(k,p) &= -\mu |k| \frac{\alpha_s(1-\alpha_s^2)}{1-\alpha_s\alpha_d}, \\ G_{12}(k,p) &= ik\mu(2 - \frac{1-\alpha_s^2}{1-\alpha_s\alpha_d}) = -G_{21}(k,p), \end{aligned} \tag{8}$$

and

$$\alpha_s = \sqrt{1 + p^2 / k^2 c_s^2},$$
$$\alpha_d = \sqrt{1 + p^2 / k^2 c_d^2}.$$
(9)

In the above equations, $c_s = \sqrt{\mu/\rho}$ is the shear wave speed and $c_d = \sqrt{(\lambda + 2\mu)/\rho}$ is the dilatational wave speed (Here, $\rho$ is the density of the solid). Eqn. (7) can be inverted in the form

$$\begin{Bmatrix} U_1 \\ U_2 \end{Bmatrix} = \begin{bmatrix} C_{11} & C_{12} \\ C_{21} & C_{22} \end{bmatrix} \begin{Bmatrix} T_1 \\ T_2 \end{Bmatrix},$$
(10)

where

$$C_{11}(k,p) = -\frac{1}{\mu|k|} \frac{\alpha_s(1-\alpha_s^2)}{4\alpha_s\alpha_d - (1+\alpha_s^2)^2},$$

$$C_{22}(k,p) = -\frac{1}{\mu|k|} \frac{\alpha_d(1-\alpha_s^2)}{4\alpha_s\alpha_d - (1+\alpha_s^2)^2},$$
(11)

$$C_{12}(k,p) = \frac{1}{ik\mu} \frac{2\alpha_s\alpha_d - (1+\alpha_s^2)}{4\alpha_s\alpha_d - (1+\alpha_s^2)^2} = -C_{21}(k,p).$$

Multiplying Eqn. (10) by $p$ and extracting the radiation damping term, we get

$$\begin{Bmatrix} \frac{c_s}{\mu} T_1 + pU_1 \\ \frac{c_d}{\lambda + 2\mu} T_2 + pU_2 \end{Bmatrix} = \begin{bmatrix} M_{11} & M_{12} \\ M_{21} & M_{22} \end{bmatrix} \begin{Bmatrix} T_1 \\ T_2 \end{Bmatrix},$$
(12)

where

$$M_{11}(k,p) = \frac{c_s}{\mu} - \frac{p}{\mu|k|} \frac{\alpha_s(1-\alpha_s^2)}{4\alpha_s\alpha_d - (1+\alpha_s^2)^2},$$

$$M_{22}(k,p) = \frac{c_d}{\lambda+2\mu} - \frac{p}{\mu|k|} \frac{\alpha_d(1-\alpha_s^2)}{4\alpha_s\alpha_d - (1+\alpha_s^2)^2}, \qquad (13)$$

$$M_{12}(k,p) = \frac{p}{ik\mu} \frac{2\alpha_s\alpha_d - (1+\alpha_s^2)}{4\alpha_s\alpha_d - (1+\alpha_s^2)^2} = -M_{21}(k,p).$$

The corresponding equations for the bottom half space are obtained by replacing $U_2$ by $-U_2$ and $T_1$ by $-T_1$ and by taking all field quantities and material properties relevant to the lower half-space. Thus

$$\left\{ \begin{array}{c} \frac{c_s}{\mu}T_1 - pU_1 \\ \frac{c_d}{\lambda+2\mu}T_2 - pU_2 \end{array} \right\} = \left[ \begin{array}{cc} M_{11} & -M_{12} \\ -M_{21} & M_{22} \end{array} \right] \left\{ \begin{array}{c} T_1 \\ T_2 \end{array} \right\}. \qquad (14)$$

Adding the equations for the upper and lower half-spaces, Eqn. (12) and Eqn. (14), and invoking traction continuity, we get

$$\left\{ \begin{array}{c} \left(\frac{c_s^+}{\mu^+} + \frac{c_s^-}{\mu^-}\right)T_1 + p(U_1^+ - U_1^-) \\ \left(\frac{c_d^+}{\lambda^++2\mu^+} + \frac{c_d^-}{\lambda^-+2\mu^-}\right)T_2 + p(U_2^+ - U_2^-) \end{array} \right\} = \left[ \begin{array}{cc} M_{11}^+ + M_{11}^- & M_{12}^+ - M_{12}^- \\ M_{21}^+ - M_{21}^- & M_{22}^+ + M_{22}^- \end{array} \right] \left\{ \begin{array}{c} T_1 \\ T_2 \end{array} \right\}. \qquad (15)$$

It may noted that multiplication of the Laplace transforms of the displacement discontinuities in the second terms on the left hand side by $p$ corresponds to differentiation with respect to time of the displacement discontinuities in the time-domain. Thus, we have arrived at an elastodynamic relation between the tractions and the time derivatives of the displacement discontinuities at the interface in the Laplace-Fourier domain.

**3. Inversion of Kernels:**

In this section, we obtain the inverse Laplace transforms of the convolution kernels, $M_{ij}(k,p)$, in Eqn. (13).

**3.1. Inversion of $M_{11}(k,p)$:**

From Eqn. (13),

$$M_{11}(k,p) = \frac{c_s}{\mu}\left[1 - \frac{p}{|k|c_s}\frac{\alpha_s(1-\alpha_s^2)}{4\alpha_s\alpha_d - (1+\alpha_s^2)^2}\right]. \qquad (16)$$

Define $s = p/|k|c_s$ and $a = c_d/c_s$, so that

$$\begin{aligned}
M_{11}(k,p) &= M_{11}(s) \\
&= \frac{c_s}{\mu}\left[1 + \frac{s^3\sqrt{1+s^2}}{4\sqrt{1+s^2}\sqrt{1+s^2/a^2} - (2+s^2)^2}\right] \\
&= \frac{c_s}{\mu}s\left[\frac{1}{s} + \frac{s^2\sqrt{1+s^2}}{4\sqrt{1+s^2}\sqrt{1+s^2/a^2} - (2+s^2)^2}\right] \\
&= \frac{c_s}{\mu}s\left[\frac{1}{s} + I(s)\right].
\end{aligned} \qquad (17)$$

The last equation above defines $I(s)$. The inverse Laplace transform of $M_{11}(k,p)$ may be written as

$$M_{11}(k,t) = \frac{1}{2\pi i} \int_B M_{11}(k,p) e^{pt} dp$$
$$= c_s |k| \frac{1}{2\pi i} \int_B M_{11}(s) e^{sT} ds \qquad (18)$$

where $B$ is the Bromwich contour shown in Fig. 1 and $T = c_s |k| t$. Using Eqn. (17) in the above equation and observing that the Laplace inverse of $1/s$ is unity (a constant), we can write

$$M_{11}(k,t) = c_s |k| \frac{c_s}{\mu} \frac{dI}{dT}, \qquad (19)$$

where

$$I(T) = \frac{1}{2\pi i} \int_B I(s) e^{sT} ds$$
$$= \frac{1}{2\pi i} \int_B \frac{s^2 \sqrt{1+s^2}}{4\sqrt{1+s^2}\sqrt{1+s^2/a^2} - (2+s^2)^2} e^{sT} ds$$
$$= \frac{1}{2\pi i} \int_B \frac{s^2 \sqrt{1+s^2}(4\sqrt{1+s^2}\sqrt{1+s^2/a^2} + (2+s^2)^2)}{-s^2(s^2+y_R^2)(s^2+y_1^2)(s^2+y_2^2)} e^{sT} ds \qquad (20)$$
$$= -\frac{1}{2\pi i} \int \frac{4(1+s^2)\sqrt{1+s^2/a^2} + (2+s^2)^2 \sqrt{1+s^2}}{(s^2+y_R^2)(s^2+y_1^2)(s^2+y_2^2)} e^{sT} ds.$$

To ensure single valuedness of the integrand in the above equation, a branch cut is taken in the complex $s$-plane from $s = -ia$ to $s = ia$. The integrand has poles at $s = \pm i y_R$, where $y_R = c_R / c_s$ and $c_R$ is the Rayleigh wave speed. The branch cuts defined above and the poles at $s = \pm i y_R$ are the only singularities of the integrand in the $s$-plane. The other zeroes of the denominator at $s = \pm i y_1$ and $s = \pm i y_2$ do not constitute poles since the numerators have zeroes at the same locations. It follows from Cauchy's theorem that the contour of integration in Eqn. (20) above can be shrunk onto the branch cut. To properly treat the poles at $s = \pm i y_R$ that lie on the branch cut, we deform the contour in the vicinity of the poles into semi-circular arcs of radius $\rho$ (not to be confused with the notation for the density) as shown in Fig. 1 and take the limit $\rho \to 0$. Noting that the value of the integrand on the left bank of the branch cut is the negative of its value on the right bank (for the same value of $y$), it can be shown that as $\rho \to 0$, the contribution to the integral from the arc on the left bank of the cut exactly cancels that from the arc on right bank of the cut. Hence, as $\rho \to 0$,

$$I(T) = \frac{2}{\pi} \int_1^a \frac{4(1-y^2)\sqrt{1-y^2/a^2}}{(y^2-y_R^2)(y-y_1^2)(y-y_2^2)} \cos(yT) dy$$
$$+ \frac{2}{\pi} PV \int_0^1 \frac{4(1-y^2)\sqrt{1-y^2/a^2} + (2-y^2)\sqrt{1-y^2}}{(y^2-y_R^2)(y-y_1^2)(y-y_2^2)} \cos(yT) dy. \qquad (21)$$

In the above equation, $PV$ denotes the Cauchy principal value. As $T \to \infty$, the first integral above vanishes, according to the Riemann-Lebesgue lemma. However, the second integral has a non-vanishing long-term component. To see this, consider the principal value integral

$$J(T) = PV \int_0^1 \frac{g(y)}{y-y_R} e^{iyT} dy, \quad 0 < y_R < 1, \qquad (22)$$

where $g(y)$ have no singularities in the domain of integration. Now,

$$J(T) = J_o(T) + g(y_R) J_1(T), \qquad (23)$$

where

$$J_o(T) = \int_0^1 \frac{g(y) - g(y_R)}{y - y_R} e^{iyT} dy \text{ and } J_1(T) = PV \int_0^1 \frac{e^{iyT}}{y - y_R} dy. \qquad (24)$$

Clearly $J_o(T)$ is well-defined and $J_o(T) \to 0$ as $T \to \infty$. We can write $J_1(T)$ as

$$J_1(T) = e^{iy_R T} PV \int_0^1 \frac{e^{i(y-y_R)T}}{y - y_R} dy$$

$$= e^{iy_R T} PV \int_{-y_R}^{1-y_R} \frac{e^{iuT}}{u} du \qquad (25)$$

$$= e^{iy_R T} \left[ PV \int_{-y_R}^{1-y_R} \frac{\cos(uT)}{u} du + i \int_{-y_R}^{1-y_R} \frac{\sin(uT)}{u} du \right].$$

Observing that $PV \int_{-X}^{X} (\cos(uT)/u) du = 0$ for all $X$, we can write

$$J_1(T) = e^{iy_R T} \left[ -\int_{1-y_R}^{y_R} \frac{\cos(uT)}{u} du + i \int_{-y_R}^{1-y_R} \frac{\sin(uT)}{u} du \right]$$

$$= e^{iy_R T} \left[ -\int_{(1-y_R)T}^{y_R T} \frac{\cos\theta}{\theta} d\theta + i \int_{-y_R T}^{(1-y_R)T} \frac{\sin\theta}{\theta} d\theta \right]. \qquad (26)$$

Using these results, we can rewrite Eqn. (20) as

$$I(T) = \int_0^1 \frac{g(y) - g(y_R)}{y - y_R} \cos(yT) dy - g(y_R) \left[ \cos(y_R T) \int_{(1-y_R)T}^{y_R T} \frac{\cos\theta}{\theta} d\theta + \sin y_R T \int_{-y_R T}^{(1-y_R)T} \frac{\sin\theta}{\theta} d\theta \right]$$

$$+ \frac{2}{\pi} \int_1^a \frac{4(1-y^2)\sqrt{1-y^2/a^2}}{(y^2 - y_R^2)(y^2 - y_1^2)(y^2 - y_2^2)} \cos(yT) dy, \qquad (27)$$

where

$$g(y) = \frac{2}{\pi} \frac{4(1-y^2)\sqrt{1-y^2/a^2} + (2-y^2)^2 \sqrt{1-y^2}}{(y + y_R)(y^2 - y_1^2)(y^2 - y_2^2)}. \qquad (28)$$

Now,

$$\frac{dI}{dT} = -\int_0^1 \frac{g(y)-g(y_R)}{y-y_R} y \sin yT \, dy$$

$$+g(y_R)\left[ y_R \sin(y_R T) \int_{(1-y_R)T}^{y_R T} \frac{\cos\theta}{\theta} d\theta - y_R \cos(y_R T) \int_{-y_R T}^{(1-y_R)T} \frac{\sin\theta}{\theta} d\theta \right] \quad (29)$$

$$+g(y_R)\frac{\cos T - 1}{T} - \frac{2}{\pi}\int_1^a \frac{4(1-y^2)\sqrt{1-y^2/a^2}}{(y^2-y_R^2)(y^2-y_1^2)(y^2-y_2^2)} y \sin(yT) \, dy.$$

Using this in Eqn. (19) and writing $|k|c_s t$ for $T$, we get,

$$M_{11}(k,t) = c_s |k| \frac{c_s}{\mu}$$

$$\left\{ \begin{array}{l} -\int_0^1 \frac{g(y)-g(y_R)}{y-y_R} y \sin(y|k|c_s t) dy \\[6pt] +g(y_R)\left[ y_R \sin(|k|c_R t) \int_{(|k|c_s t-|k|c_R t)}^{|k|c_R t} \frac{\cos\theta}{\theta} d\theta - y_R \cos(|k|c_R t) \int_{-|k|c_R t}^{(|k|c_s t-|k|c_R t)} \frac{\sin\theta}{\theta} d\theta \right] \\[6pt] +g(y_R)\frac{\cos(|k|c_s t)-1}{|k|c_s t} \\[6pt] -\frac{2}{\pi}\int_1^a \frac{4(1-y^2)\sqrt{1-y^2/a^2}}{(y^2-y_R^2)(y^2-y_1^2)(y^2-y_2^2)} y \sin(y|k|c_s t) dy \end{array} \right\}.$$

(30)

The integrals in the above equation are not amenable to analytical solution and are evaluated numerically.

The long-time response can be found in closed form are

$$M_{11}(k,t\to\infty) = -c_s |k| \frac{c_s}{\mu} \frac{4(1-y_R^2)\sqrt{1-y_R^2/a^2} + (2-y_R^2)^2\sqrt{1-y_R^2}}{(y_R^2-y_1^2)(y_R^2-y_2^2)} \cos(|k|c_R t). \quad (31)$$

**3.2. Inversion of $M_{22}(k,p)$:**

The inversion of $M_{22}(k,p)$ follows a similar procedure to that for $M_{11}(k,p)$. The integration contour is the same as that shown in Fig. 1. We finally obtain

$$M_{22}(k,t) = c_s |k| \frac{c_s}{\mu} \left\{ \begin{array}{l} -\int_0^1 \frac{f(y)-f(y_R)}{y-y_R} y\sin(y|k|c_s t) dy \\ + f(y_R)\left[ y_R \sin(|k|c_R t) \int_{(|k|c_s t - |k|c_R t)}^{|k|c_R t} \frac{\cos\theta}{\theta} d\theta - y_R \cos(|k|c_R t) \int_{-|k|c_R t}^{(|k|c_s t - |k|c_R t)} \frac{\sin\theta}{\theta} d\theta \right] \\ + f(y_R) \frac{\cos(|k|c_s t)-1}{|k|c_s t} \\ - \frac{2}{\pi} \int_1^a \frac{(2-y^2)^2 \sqrt{1-y^2/a^2}}{(y^2-y_R^2)(y^2-y_1^2)(y^2-y_2^2)} y\sin(y|k|c_s t) dy \end{array} \right\}, \quad (32)$$

where

$$f(y) = \frac{2}{\pi} \frac{4(1-y^2/a^2)\sqrt{1-y^2} + (2-y^2)^2 \sqrt{1-y^2/a^2}}{(y+y_R)(y^2-y_1^2)(y^2-y_2^2)}. \quad (33)$$

The long-time response is

$$M_{22}(k,t\to\infty) = -c_s |k| \frac{c_s}{\mu} \frac{4(1-y_R^2/a^2)\sqrt{1-y_R^2} + (2-y_R^2)^2 \sqrt{1-y_R^2/a^2}}{(y_R^2-y_1^2)(y_R^2-y_2^2)} \cos(|k|c_R t). \quad (34)$$

### 3.3. Inversion of $M_{12}(k,p)$:

From Eqn. (13), we have

$$M_{12}(k,p) = -i\,\mathrm{sgn}(k) \frac{c_s}{\mu} \frac{p}{|k|c_s} \frac{2\alpha_s \alpha_d - (1+\alpha_s^2)}{4\alpha_s \alpha_d - (1+\alpha_s^2)^2}. \quad (35)$$

Writing $s = p/|k|c_s$ and $a = c_d/c_s$ as before,

$$M_{12}(k,p) = i\,\text{sgn}(k)\frac{c_s}{\mu}s\left[\frac{\left[8s^2/a^2 + 8/a^2 - s^4 - 4 - 6s^2 + 2\sqrt{1+s^2}\sqrt{1+s^2/a^2}(2+s^2)\right]}{(s^2+y_R^2)(s^2+y_1^2)(s^2+y_2^2)}\right]$$

$$= i\,\text{sgn}(k)\frac{c_s}{\mu}sI(s). \tag{36}$$

The last equation above defines $I(s)$. It follows that

$$M_{12}(k,t) = i\,\text{sgn}(k)|k|c_s\frac{c_s}{\mu}\frac{1}{2\pi i}\int_B sI(s)e^{sT}\,ds = ikc_s\frac{c_s}{\mu}\frac{dI(T)}{dT}, \tag{37}$$

where $B$ is the Bromwich contour shown in Fig. 2. On performing the inversion, we get

$$I(T) = \frac{\left[-8y_R^2/a^2 + 8/a^2 - y_R^4 - 4 + 6y_R^2 + 2\sqrt{1-y_R^2}\sqrt{1-y_R^2/a^2}(2-y_R^2)\right]}{(y_1^2-y_R^2)(y_2^2-y_R^2)}\frac{\sin(y_R T)}{y_R}$$

$$-\frac{2}{\pi}\int_1^a \frac{2\sqrt{y^2-1}\sqrt{1-y^2/a^2}(2-y^2)}{(y_R^2-y^2)(y_1^2-y^2)(y_2^2-y^2)}\sin(yT)\,dy. \tag{38}$$

This gives

$$M_{12}(k,t) = ikc_s\frac{c_s}{\mu}$$

$$\left\{\begin{array}{l}-\dfrac{2}{\pi}\displaystyle\int_1^a \dfrac{2y\sqrt{y^2-1}\sqrt{1-y^2/a^2}(2-y^2)}{(y_R^2-y^2)(y_1^2-y^2)(y_2^2-y^2)}\cos(y|k|c_s t)\,dy \\[1em] +\left(\dfrac{\left[-8y_R^2/a^2 + 8/a^2 - y_R^4 - 4 + 6y_R^2 + 2\sqrt{1-y_R^2}\sqrt{1-y_R^2/a^2}(2-y_R^2)\right]}{(y_1^2-y_R^2)(y_2^2-y_R^2)}\right)\cos(|k|c_R t)\end{array}\right\}. \tag{39}$$

The last term in the above equation clearly gives the long-time response.

The kernels $M_{11}(k,t)$, $M_{22}(k,t)$ and $M_{12}(k,t)$ are plotted in Fig. 3 for a solid with a Poisson's ratio of 0.25. The long-time response of the three kernels is in agreement with the analytically derived expressions given in Eqns. (31), (34) and (39).

## 4. Validation of Kernels:

The convolution kernels obtained in the previous section are validated by studying two problems: (1) a modal analysis and (2) studying the response to an impulsive line load on an elastic half-space. Analytical solutions are known for these problems and good agreement between the analytical solutions and the numerical results validates the kernels.

### 4.1. Modal Analysis:

The modal analysis is performed for an interface between identical elastic solids loaded in tension. We assume that a cohesive law is operative at the interface. Under the action of a remote tensile stress, $\sigma_o$, a steady state opening, $\delta_{2o}$, develops at the interface. Due to symmetry, there is no shear stress at the interface. We obtain both the analytical and numerical solution to an opening velocity perturbation at the interface between the solids in a single Fourier mode of the form

$$\dot{\delta}_2(x_1,t) = H(t)\exp(ikx_1), \qquad (40)$$

where $H(t)$ is the Heaviside step function. To avoid interpenetration of the solids, we assume the magnitude of the perturbation to be small, $|\delta_2| \ll \delta_{2o}$. From Eqn. (15), the relation between the opening displacement perturbation $D_2$ and the normal stress perturbation $T_2$ is

$$T_2 = -\frac{\mu|k|}{2}\frac{4\alpha_s\alpha_d - (1+\alpha_s^2)^2}{\alpha_d(1-\alpha_s^2)}D_2. \tag{41}$$

The instantaneous normal stress change when the velocity perturbation of the form Eqn. (40) is applied is $-(\lambda + 2\mu)/2c_d$. Non-dimensionalizing $T_2$ in Eqn. (41) by this quantity and substituting $D_2 = 1/p^2$, we get

$$r = \frac{T_2}{-(\lambda+2\mu)/2c_d} = \frac{c_d \mu |k|}{(\lambda+2\mu)p^2}\frac{4\alpha_s\alpha_d - (1+\alpha_s^2)^2}{\alpha_d(-s^2)} = -\frac{1}{c_d|k|}\left[\frac{4\alpha_s}{s^4} - \frac{(1+\alpha_s^2)^2}{\alpha_d s^4}\right]. \tag{42}$$

Performing the inverse Laplace transform explicitly, we obtain

$$r(T) = \frac{1}{a}\left\{aJ_0(aT) + \int_0^T (aJ_0(a(T-T')) - J_0(T-T'))(\frac{2T'^3}{3} + 4T').dT'\right\}, \tag{43}$$

where $T = |k|c_s t$, $a = c_d/c_s$ and $J_0$ is the Bessel function of the first kind of order zero. The same quantity $r(T)$ can be obtained numerically using the convolution kernel $M_{22}$ defined earlier. From Eqn. (15), we have

$$\frac{2c_d}{\lambda+2\mu}T_2 + pD_2 = 2M_{22}T_2. \tag{44}$$

Introducing $r = -2c_d T_2/(\lambda+2\mu)$, this become

$$-r + pD_2 = -M_{22}(\lambda+2\mu)r/c_d \tag{45}$$

Taking inverse Laplace transform, the second term on the left give $\dot{D}_2$. Since we have $\dot{\delta}_2(x_1,t) = H(t)\exp(ikx_1)$, by comparison $\dot{D}_2 = H(t)$. For $t \geq 0$, this is unity. The term on the right-hand side gives a convolution integral on inverting. Putting everything together, we get

$$r(t) = 1 + 2\frac{\lambda+2\mu}{2c_d}\int_0^t M_{22}(t-t')r(t')dt'. \tag{46}$$

Observing from Eqn. (32) that $M_{22}(t)$ has the form $M_{22}(t) = \frac{|k|c_s^2}{\mu} m_{22}(T)$, we get from Eqn. (46)

$$r(T) = 1 + \frac{c_d}{c_s} \int_0^T m_{22}(T-T')r(T')dT', \qquad (47)$$

which is a Volterra equation of the second kind. Solving that Volterra equation numerically, we can determine $r(T)$. The numerical solution is compared with the analytical solution, Eqn. (43), in Fig. 4 when both solids have a Poisson's ratio of 0.25. The numerical solution of Eqn. (47) depends on a discretization parameter, $\gamma = \Delta T$. Three numerical computations are shown with values of $\gamma$ being 2, 1 and 0.1. It is seen that the result with $\gamma = 0.1$ agrees well with the analytical solution.

### 4.2. Impulsive Line-Load on a Half-Space:

A half-space is loaded by a horizontal impulsive line-load and the displacements at a point on the free surface are determined. The impulsive line load is given by

$$\tau_1^0 = P\delta(x_1)\delta(t), \quad \tau_2^0 = 0, \qquad (48)$$

where $P$ is the magnitude of the load. The free surface conditions translate to $\tau_1 = 0, \tau_2 = 0$. Substituting into Eqn. (12), we obtain the elastodynamic equations for each Fourier mode as

$$\frac{c_s}{\mu} \dot{T}_1 + pU_1 = M_{11}T_1,$$
$$pU_2 = M_{21}T_1. \qquad (49)$$

The free surface is discretized by a grid of 512 points. A Fast Fourier Transform operation is performed to obtain $T_1(k,t)$ for the horizontal impulsive load. The velocities in the $x_1$ and $x_2$ directions at all grid points are calculated from Eqn. (49). This involves performing the convolutions indicated in Eqn. (49) and also an inverse Fast Fourier Transform operation. The time step, $\Delta t$, is chosen in terms of the time taken by the shear wave to traverse one grid spacing as $\Delta t = \beta \Delta x_1 / c_s$. The value of $\beta$ is chosen so that the numerical convolution for even the highest mode is accurately performed. In the modal analysis of Section 4.1, it was seen that $\gamma = 0.1$ is required for accurately solving the modal equation. Following Morrissey and Geubelle (1997), it may be shown that this translates to the condition that $\beta < 0.1/\pi = 0.031$. Therefore, the value of $\beta$ is taken to be 0.01. The displacements are calculated by integrating the velocities. The displacements obtained at a point $x_1 = X$ which is 100 grid-points away from the load-point are plotted in Figs. 5 and 6. They are compared with the analytical solutions given in Eringen and Suhubi (1975). Poisson's ratio is taken to be 0.25. The solution for $u_1$ has a propagating singularity and the solution for $u_2$ has a propagating delta function, both with speed of the Rayleigh wave. These are seen to well captured in the numerical results. Oscillations are seen in the solutions, especially in the vicinity of singularities and the delta function. These may be ascribed to the inability of a finite Fourier sum to adequately represent a singularity or delta function. However, all wave arrivals and their magnitudes are correctly captured in the numerical solution. This validates the kernels $M_{11}$ and $M_{12}$. Similar calculations are performed for a vertical impulse on an elastic half-space. The procedure now involves the kernels $M_{22}$ and $M_{21}$. The vertical displacement at the free surface due to a vertical impulse is shown in Fig. 7 and is compared with the analytical solution. Since the agreement with the analytical solution is excellent, this validates the kernel $M_{22}$. (Since $M_{21} = -M_{12}$, the validation already done for

$M_{12}$ makes a separate calculation of the horizontal displacement due to a vertical impulse redundant.)

## 5. Discussion:

The convolution kernels in a spectral formulation of the boundary integral equations for an interface between dissimilar elastic half-spaces have been derived. When combined with a constitutive law for the interface, the spectral formulation can be used for simulating dynamic ruptures at the interface between two elastic solids. We illustrate below how the method proposed here compares with that of Breitenfeld and Geubelle (1998).

In a numerical simulation, the spectral components of the displacement discontinuities and tractions are evaluated as

$$\begin{aligned}\delta_j(x_1,t) &= D_j(k,t)\exp(ikx_1), \\ \tau_j(x_1,t) &= T_j(k,t)\exp(ikx_1),\end{aligned} \quad (50)$$

where $j=1,2$. The elastodynamic relations between $T_j(k,t)$ and $D_j(k,t)$ are obtained by taking the inverse Laplace transform of Eqn. (15) as

$$\begin{aligned}&\left\{\begin{array}{l}\left(\dfrac{c_s^+}{\mu^+}+\dfrac{c_s^-}{\mu^-}\right)T_1(k,t)+\dfrac{\partial D_1(k,t)}{\partial t} \\ \left(\dfrac{c_d^+}{\lambda^++2\mu^+}+\dfrac{c_d^-}{\lambda^-+2\mu^-}\right)T_2(k,t)+\dfrac{\partial D_2(k,t)}{\partial t}\end{array}\right\} = \\ &\int_0^t \begin{bmatrix} M_{11}^+(k,t')+M_{11}^-(k,t') & M_{12}^+(k,t')-M_{12}^-(k,t') \\ M_{21}^+(k,t')-M_{21}^-(k,t') & M_{22}^+(k,t')+M_{22}^-(k,t') \end{bmatrix}\begin{Bmatrix} T_1(k,t-t') \\ T_2(k,t-t') \end{Bmatrix}dt'.\end{aligned} \quad (51)$$

This elastodynamic relation needs to be combined with a constitutive law for the interface, which in general could be of the general form

$$\tau_j(x_1,t) = \tau_j(\delta_1,\delta_2,\dot{\delta}_1,\dot{\delta}_2,\theta_1,\theta_2,...,\theta_n), j=1,2, \qquad (52)$$

where $\theta_j$, $j=1,...,n$ are state variables which characterize the memory dependence of the shear and normal stresses on the slip and opening. It may be noted that the solution of the bi-material rupture problem in its most general form involves the evaluation of four convolution integrals in Eqn. (51) at each time instant. This may be contrasted with the approach of Breitenfeld and Geubelle (1998). They developed a so-called "independent formulation" where the spectral elastodynamic relations are developed separately for each half space. For the upper half-space, they write

$$\begin{Bmatrix} T_1(k,t) + \dfrac{\mu^+}{c_s^+}\dfrac{\partial U_1^+(k,t)}{\partial t} \\ T_2(k,t) + \dfrac{\lambda^+ + 2\mu^+}{c_d^+}\dfrac{\partial U_2^+(k,t)}{\partial t} \end{Bmatrix} = \int_0^t \begin{bmatrix} H_{11}^+(k,t') & H_{12}^+(k,t') \\ H_{21}^+(k,t') & H_{22}^+(k,t') \end{bmatrix} \begin{Bmatrix} U_1^+(k,t-t') \\ U_2^+(k,t-t') \end{Bmatrix} dt' \qquad (53)$$

and for the lower half-space, they have

$$\begin{Bmatrix} T_1(k,t) - \dfrac{\mu^-}{c_s^-}\dfrac{\partial U_1^-(k,t)}{\partial t} \\ T_2(k,t) - \dfrac{\lambda^- + 2\mu^-}{c_d^-}\dfrac{\partial U_2^-(k,t)}{\partial t} \end{Bmatrix} = \int_0^t \begin{bmatrix} H_{11}^-(k,t') & H_{12}^-(k,t') \\ H_{21}^-(k,t') & H_{22}^-(k,t') \end{bmatrix} \begin{Bmatrix} U_1^-(k,t-t') \\ U_2^-(k,t-t') \end{Bmatrix} dt'. \qquad (54)$$

The convolution kernels $H_{ij}^\pm(k,t)$, $i,j=1,2$ were derived by Breitenfeld and Geubelle (1998). Now, Eqns. (53) and (54) need to be solved together in conjunction with the constitutive law

Eqn. (52). As is seen from Eqns. (49) and (50), this method involves the evaluation of eight convolution integrals at each instant of time. Since evaluation of the convolution integrals constitutes the major computational effort in such problems, a substantial saving in the computation time may be expected for the method proposed here in comparison to the method of Breitenfeld and Geubelle (1998). Breitenfeld and Geubelle (1998) also appear to have developed a "combined formulation" where the convolutions are done on the history of interfacial displacement discontinuities for the bi-material problem. The number of convolutions reduce to four in that case, but the authors mention that the method was less stable than the "independent formulation" and do not report any results with the "combined formulation".

## 6. Conclusions:

A spectral formulation of the boundary integral equation method applied to an interface between dissimilar elastic half-spaces has been proposed. The method involves performing elastodynamic convolutions over the history of tractions at the interface between the half-spaces. Prior methods have involved convolutions over the history of displacements or displacement discontinuities at the interface. The convolution kernels are derived in this paper and shown to have a simple form. The convolution kernels are validated by comparing numerical solutions to some model problems obtained using the kernels with known analytical solutions. The formulation presented here can be used for numerical simulation of spontaneous dynamic rupture propagation at an interface between dissimilar elastic half-spaces.

## 7. Acknowledgement:

This work was supported by Department of Science and Technology, India through Science and Engineering Research Board project SR/S3/MERC/0076/2012. Discussions with Prof. J. R. Rice of Harvard University and Mr. G. Balaji of SRM University are acknowledged.

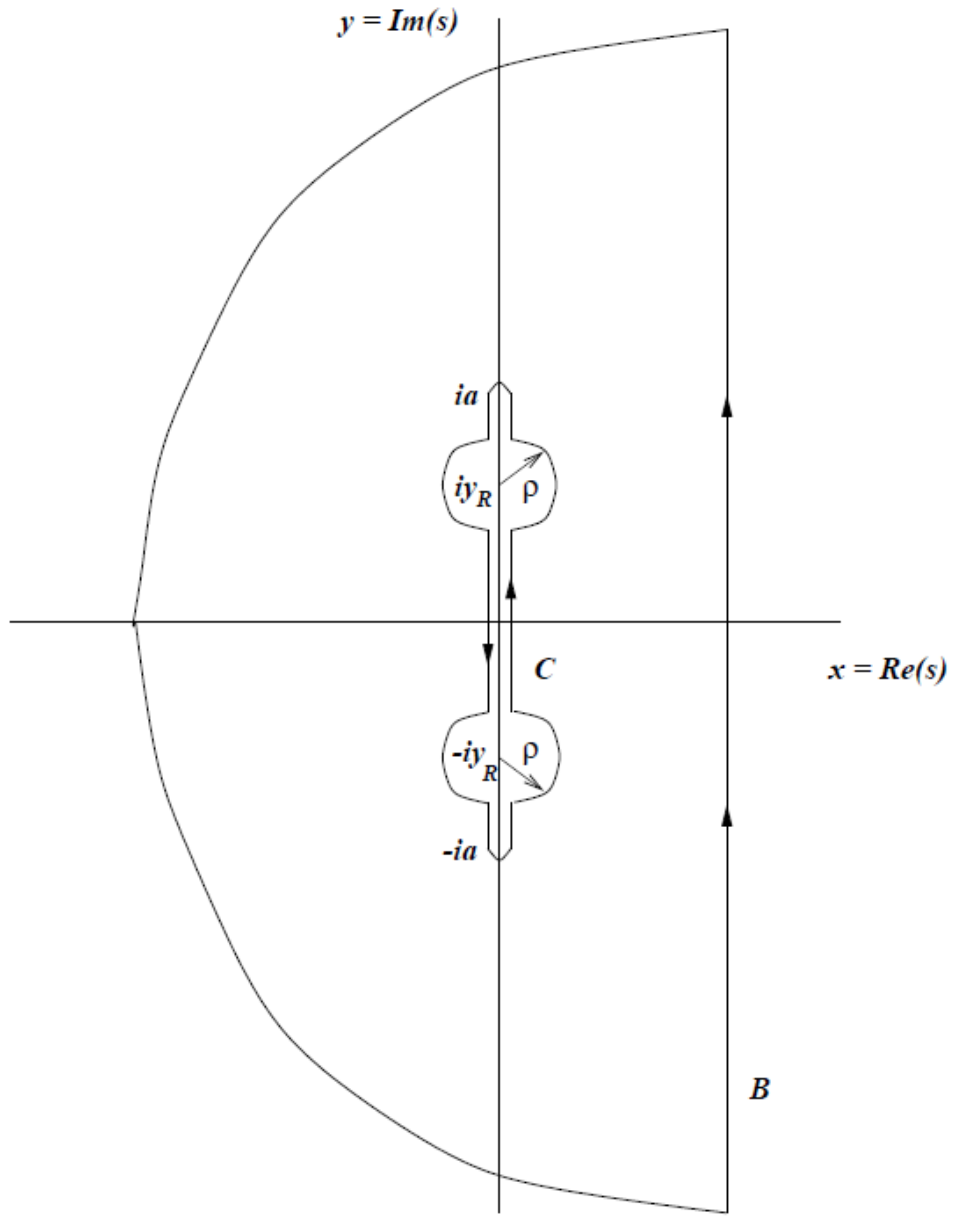

**Fig. 1** Integration contour for inverting $M_{11}(k, p)$ and $M_{22}(k, p)$

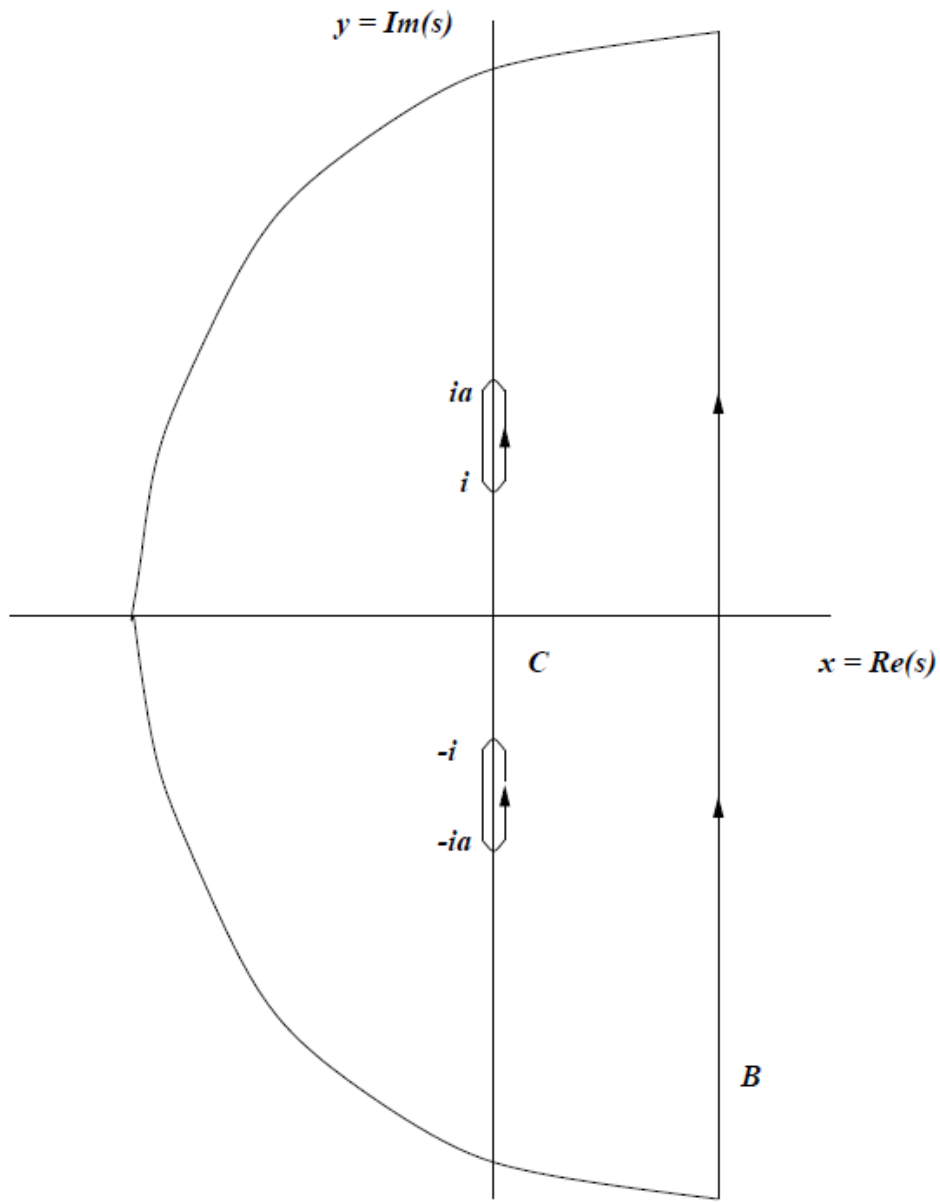

**Fig. 2** Integration contour for inverting $M_{12}(k, p)$

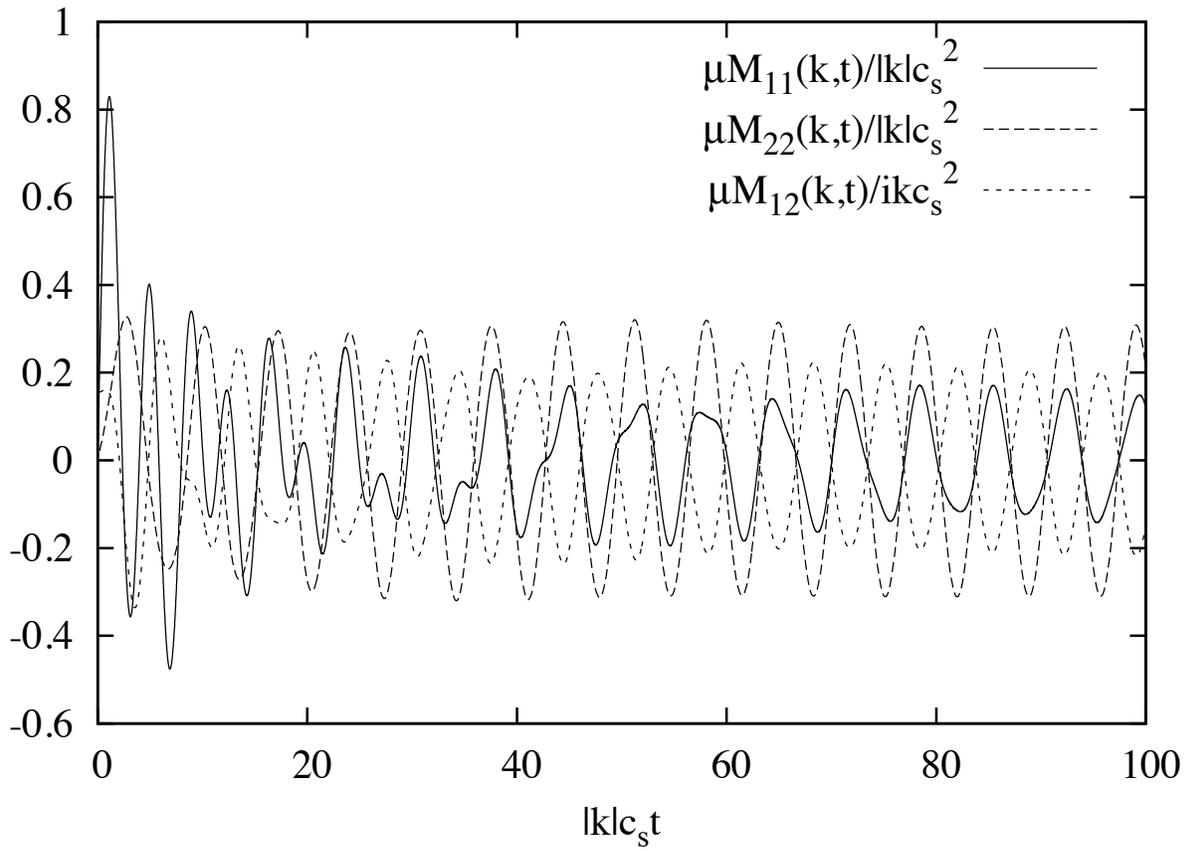

**Fig. 3** Plots of convolution kernels $M_{ij}(k,t)$ for an elastic solid with Poisson's ratio 0.25

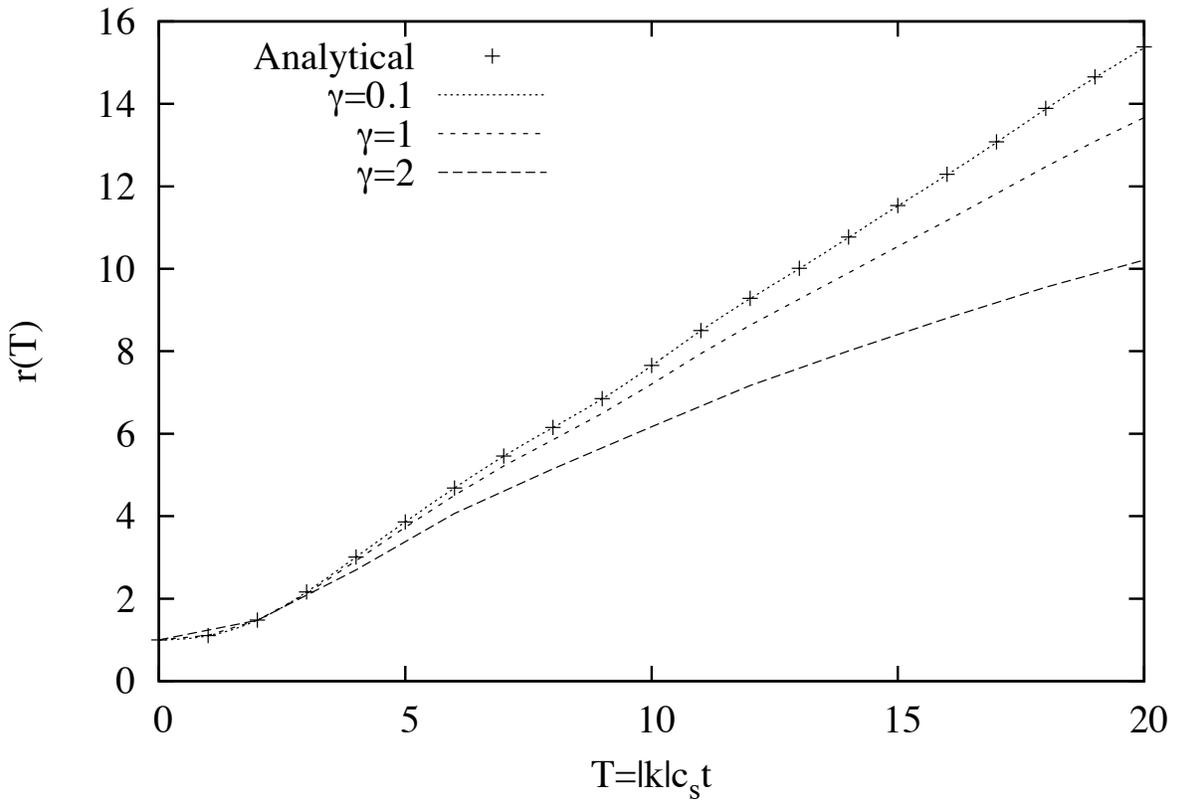

**Fig. 4** Time response of normal stress to an interfacial opening velocity perturbation in a single Fourier mode obtained for three values of the discretization parameter $\gamma$ compared with the analytical solution

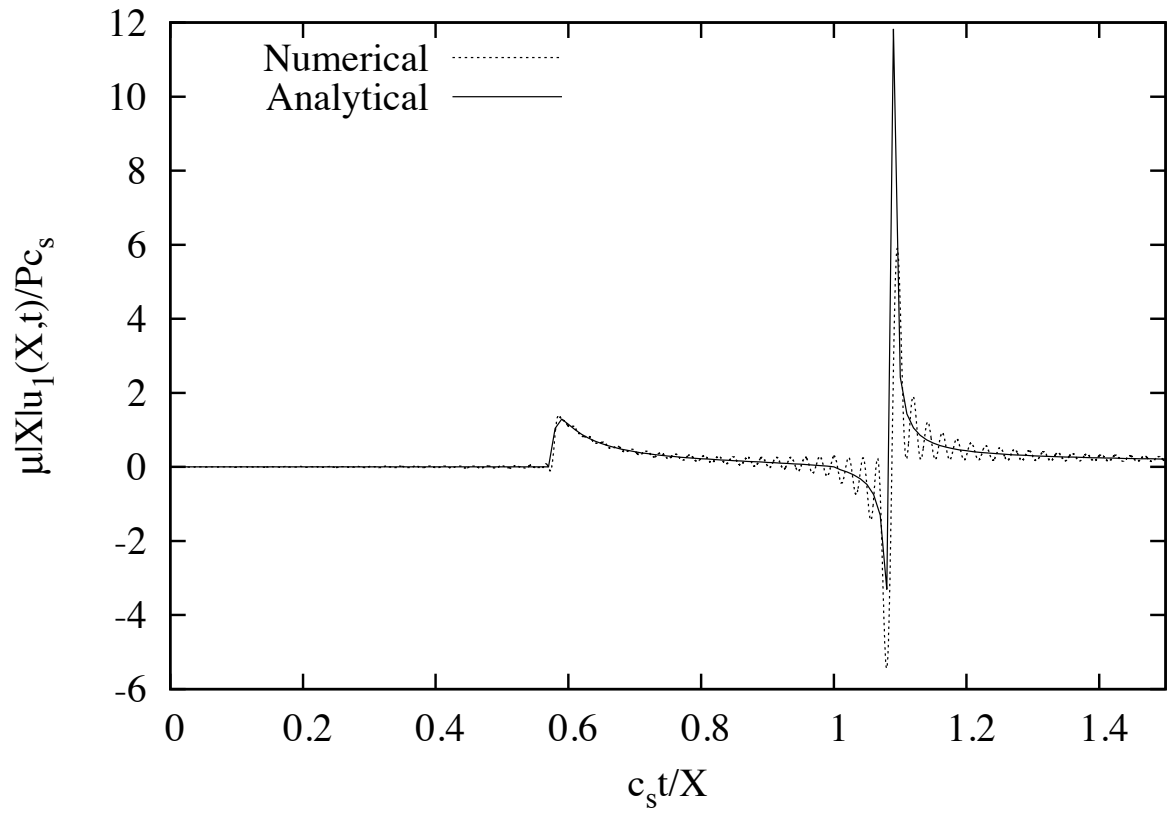

**Fig. 5** Time response of horizontal displacement due to a horizontal impulse on an elastic half-space compared with the analytical solution

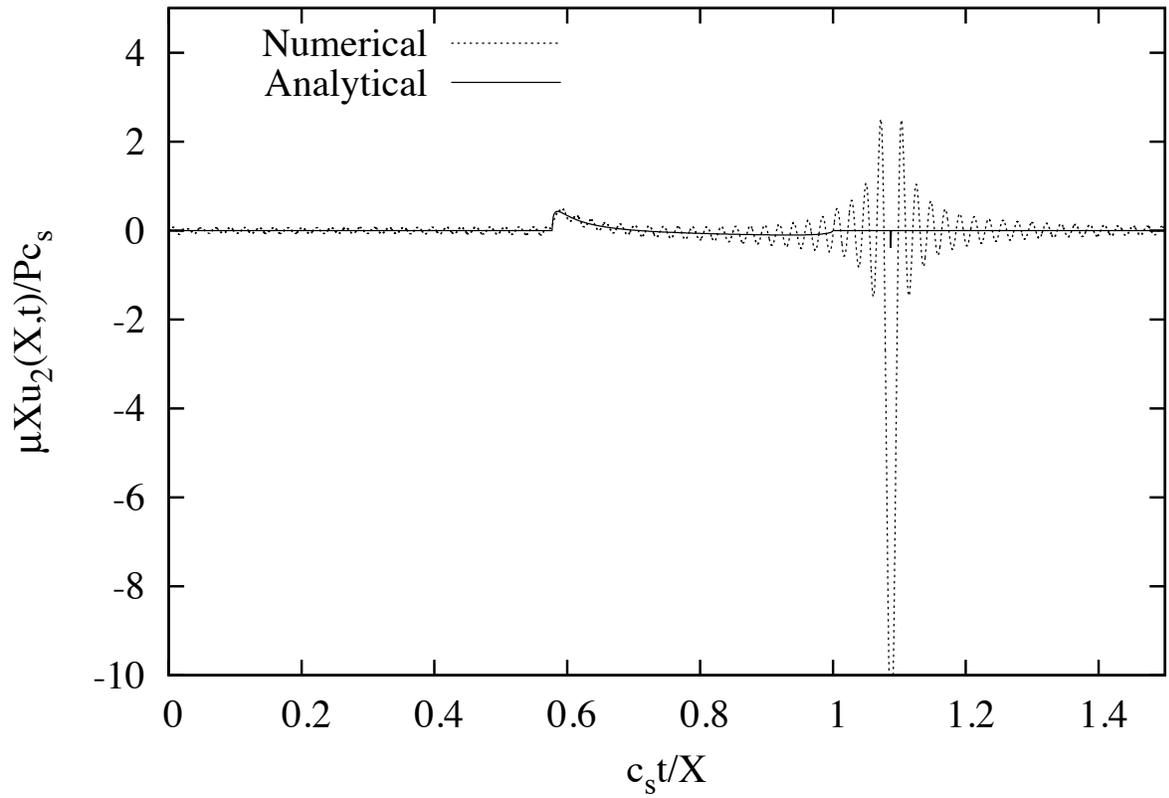

**Fig. 6** Time response of vertical displacement due to a horizontal impulse on an elastic half-space compared with the analytical solution

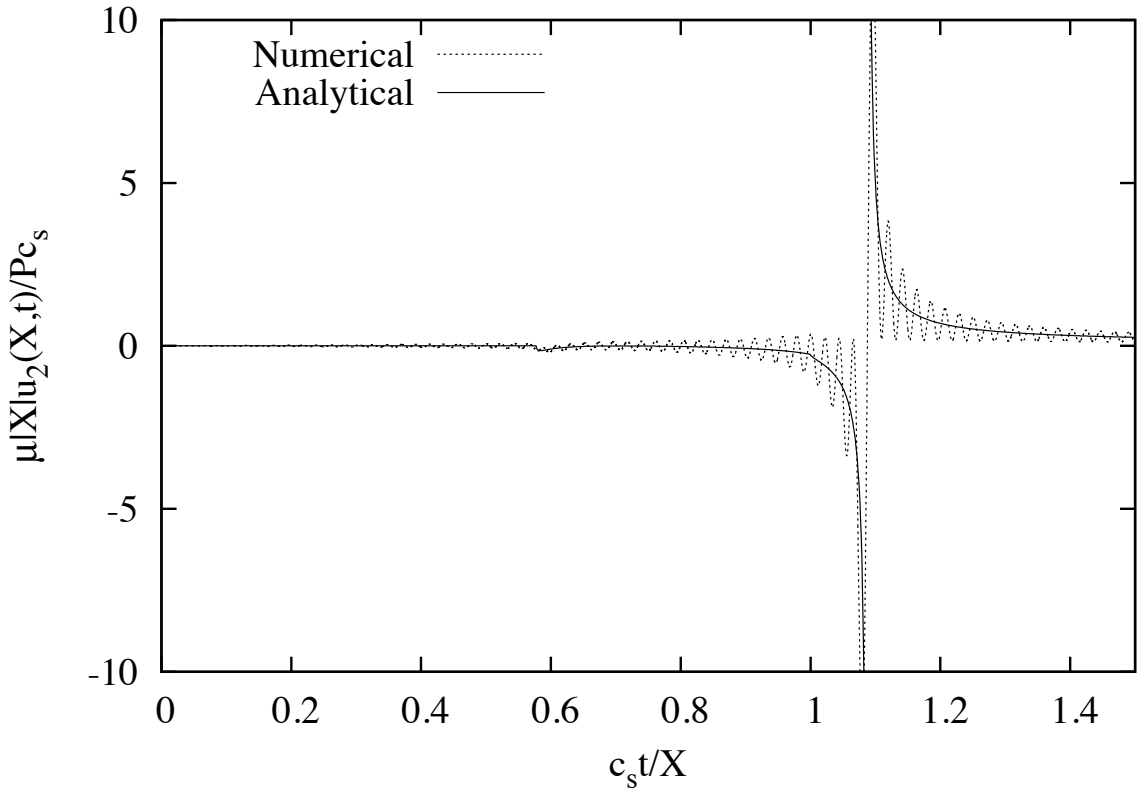

**Fig. 7** Time response of vertical displacement due to a vertical impulse on an elastic half-space compared with the analytical solution

**VITAE**

K. Ranjith is presently a Research Professor at SRM Research Institute, SRM University, India. He holds a Ph.D. degree in Engineering Sciences from Harvard University, USA, and a Master's degree in Mechanical Engineering from the Indian Institute of Science, Bangalore, India. His post-doctoral training is from the Max Planck Institute for Metals Research, Stuttgart, Germany. Previously, he has held faculty appointments at the Indian Institute of Technology, Bombay and at Amrita University, Coimbatore.

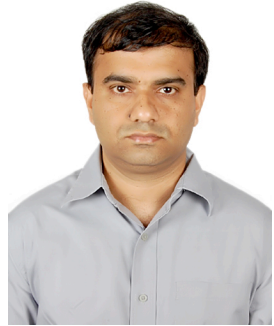

Photo of K. Ranjith